\documentclass[pdflatex,sn-mathphys-num]{sn-jnl}
\usepackage{graphicx}%
\usepackage{multirow}%
\usepackage{amsmath,amssymb,amsfonts}%
\usepackage{amsthm}%
\usepackage{mathrsfs}%
\usepackage[title]{appendix}%
\usepackage{xcolor}%
\usepackage{textcomp}%
\usepackage{manyfoot}%
\usepackage{booktabs}%
\usepackage{algorithm}%
\usepackage{algorithmicx}%
\usepackage{algpseudocode}%
\usepackage{listings}%

\usepackage{braket}

\theoremstyle{thmstyleone}%
\theoremstyle{thmstyletwo}%

\theoremstyle{thmstylethree}%

\begin{document}

\title[]{Distant Entanglement Generation between Magnon and Superconducting Qubits in Magnon-Mediated Hybrid Systems}

\author*[1]{\fnm{Guosen} \sur{Liu}}\email{lgs0866@dlmu.edu.cn}
\author[1]{\fnm{Pei} \sur{Pei}}\email{peipei@dlmu.edu.cn}

\affil[1]{\orgdiv{College of Science}, \orgname{Dalian Maritime University}, \orgaddress{\city{Dalian}, \postcode{116026}, \country{China}}}

\abstract{We propose an efficient two-stage protocol for generating distant entanglement in a magnon-mediated hybrid quantum system, where magnons serve dual roles as both interaction mediators and qubits. This integrated design reduces the physical component count while leveraging the inherent advantages of magnons, such as their strong coupling via magnetic dipole interactions, low dissipation, and high integrability. In our setup, a superconducting resonator interfaces between a local superconducting qubit (SQ) and a local magnonic system (QM1), which is waveguide-coupled to a remote magnonic system (QM2). The protocol comprises two stages: (i) deterministic Bell-state generation between the SQ and QM1 using shortcuts to adiabaticity, and (ii) coherent state transfer to QM2 via engineered Hamiltonian dynamics. This adiabatic characteristic enhances robustness against environmental dissipation. Numerical simulations under realistic noise conditions confirm strong resilience to decoherence, achieving fidelity $F > 0.90$ and negativity $\mathcal{N}_2 > 0.40$. These results establish the protocol as a scalable and practical building block for distributed quantum networks.}


\maketitle

\section{Introduction}\label{sec1}
The foundation of quantum information science lies in the controlled generation and coherent manipulation of entanglement, which are essential for implementing advanced quantum technologies such as quantum computing\cite{A4,A5}, quantum metrology\cite{A2,A3}, and quantum communication\cite{A6,A7}. 
Quantum entanglement is a non-classical correlation that grants quantum systems the ability to surpass the limitations of classical physics\cite{A1,A8,Y1}. 
However, traditional entanglement generation methods, predominantly relying on slow adiabatic evolution, are not only time-consuming but also highly susceptible to environmental noise, resulting in limited efficiency and reliability for practical applications. 
To address these limitations, alternative strategies including shortcuts to adiabaticity (STA) schemes\cite{STA1,STA2,STA3} and hybrid quantum systems\cite{Y2,B13,B14,H1,H3} have been proposed to overcome these challenges.

The practicality of quantum entanglement in quantum information processing is enhanced through leveraging the complementary strengths of different physical platforms.
For instance, superconducting qubits are characterized by high-precision control and integrability, superconducting circuit resonators provide strong coupling and longer coherence times\cite{PhysRevA.109.022442}. As an emerging platform, magnons can act as efficient qubits due to their low dissipation, strong quantum coupling mediated by magnetic dipole interactions, and nonlinear interaction capabilities with diverse quantum systems\cite{Sohail2022,M2,C4,C5,C6}. Additionally, magnons can act as mediators to efficiently interface different quantum systems; for example, a magnonic device can functionally replace both the nanomechanical resonator and optical cavity required to couple a superconducting resonator to a waveguide in a hybrid system, thereby reducing component count and improving integration\cite{052421}.
The coupling between magnons and photons enables remote transmission of quantum states, supporting quantum communication networks\cite{Y3,Y31,Y32}. Integrating magnons with superconducting systems can further enhance the speed and fidelity of quantum information processing, opening new possibilities for quantum computing and communication. 
Recent advances in magnon-mediated hybrid systems have demonstrated entanglement between magnons and superconducting qubits\cite{Y4,PhysRevA.110.053710}, magnons and photons\cite{Y5}, and photon pairs\cite{Y6}, as well as two-magnon entanglement in microwave cavity systems\cite{2025magnon,zhao2022,Y7,Y8,Y9,PhysRevA.109.012611}. These developments highlight the potential of hybrid systems for efficient quantum information processing. 

Beyond the selection of physical systems, enhancing the efficiency of entanglement generation is equally crucial. Although traditional adiabatic evolution methods can generate high-fidelity entanglement, their slow evolution speeds limit practical applications. The STA method has been extensively utilized as an effective optimization strategy, achieving faster and higher-fidelity entanglement generation in diverse physical systems.
Among STA approaches, the invariant-based method has emerged as particularly effective\cite{STA3,STA21,STA22}, enabling breakthroughs in entanglement generation. 
For instance, STA has been successfully applied to entangle superconducting qubits with magnons\cite{Y11}. The invariant-based STA method optimizes driving parameters\cite{Y10}, significantly improving the efficiency and precision of the entanglement generation. Building on these advances, this work also adopts a reduced density matrix approach to simplify the treatment of multi-bosonic mode systems\cite{Y3}. While several recent studies have demonstrated remote entanglement generation in magnon-mediated hybrid systems\cite{zhao2022,PhysRevA.109.012611,PhysRevLett.127.087203,PhysRevA.109.043708,PhysRevA.104.023711}, we propose a novel STA-based scheme for hybrid quantum systems that achieves high-fidelity and distant entanglement across distinct physical platforms.

In this paper, we organize the content as follows. 
Section~\ref{sec2} introduces a five-body hybrid quantum system based on magnons, providing the physical platform for our study. Building on this system, Section~\ref{sec3} establishes a phased approach for generating local-to-remote entanglement. The feasibility of this scheme is then numerically validated in Section~\ref{sec4}, focusing on high-fidelity entanglement generation with a high entanglement measure $\mathcal{N}_2$. Finally, Section~\ref{sec5} concludes the work, discusses potential applications, and outlines future research directions.

\section{Magnon-Mediated Hybrid Systems}\label{sec2}
\begin{figure}[h]
\centering
\includegraphics[width=0.7\textwidth]{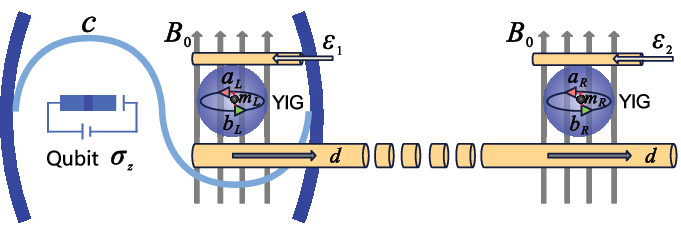}
\caption{Schematic diagram of a five-part hybrid quantum system based on magnons.}\label{Fig1}
\end{figure}

\begin{figure}[h]
\centering
\includegraphics[width=0.7\textwidth]{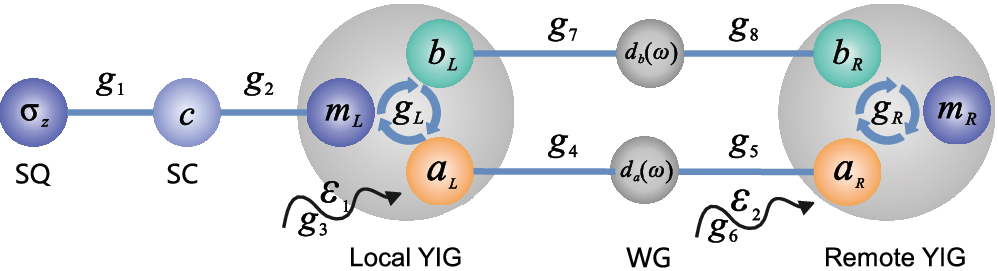}
\caption{Coupling schematic of a five-part hybrid quantum system based on magnons. The SC mode ($c$) and WG mode ($d_{a(b)}(\omega)$ originate from quantized electromagnetic fields in the microwave cavity and waveguide, respectively. The magnon modes ($m_{L}$ and $m_{R}$) are spin-wave excitations in YIG spheres, and their frequencies can be tunable through the external field $B_0$. The whispering gallery modes (WGMs) in each YIG sphere comprise low-energy photon modes ($a_{L}$ and $a_{R}$) and high-energy photon modes ($b_{L}$ and $b_{R}$), arising from the confined photonic states owing to the material's high refractive index and spherical geometry.}\label{Fig2}
\end{figure}

The proposed hybrid quantum system (Figure~\ref{Fig1}) integrates five components to enable coherent information transfer between superconducting and magnonic platforms: (i) a superconducting charge qubit (SQ)--a Josephson-junction-based quantum system that encodes information in Cooper pair charge states on a superconducting island, typically controlled by gate voltage and operating at cryogenic temperatures\cite{H2,H3}; (ii) a superconducting circuit resonator (SC) composed of two parallel ground planes with a central superconducting wire sandwiched between them, serving as a cavity\cite{H2,H3}; (iii) a local quantum magnonic system (QM1) implemented in a Yttrium Iron Garnet (YIG) sphere supporting collective spin-wave excitations\cite{Y2,H3}; (iv) a remote quantum magnonic system (QM2)  implemented in another YIG sphere; and (v) a waveguide (WG) mediating coherent coupling between QM1 and QM2. The SQ and QM1 are integrated on the SC platform, where the SQ is capacitively coupled to the SC while the SC interacts with QM1 via magnetic dipole coupling\cite{Y2,H1}. The WG mediates coherent coupling between the two quantum magnonic systems (QM1 and QM2) through the interaction of waveguide photons with whispering gallery modes (WGMs) in the YIG sphere\cite{M2,Y2}, as detailed in Figure~\ref{Fig2}.

The SQ is modeled as a two-level system, where the ground state and the excited state are denoted as $\ket{\tilde{g}}$ and $\ket{\tilde{e}}$. 
The quantum states of the superconducting circuit resonator mode and the WGMs in the YIG sphere are described by the Fock photon states $\ket{0}_{c,a,b}$ and $\ket{1}_{c,a,b}$. The quantum states of the magnons are described by the number states $\ket{0}_{m}$ and $\ket{1}_{m}$. 

In this work, the total Hamiltonian $H_{\mathrm{Total}}$ is partitioned into local and remote components, as given by:
\begin{equation}\label{eq:1}
H_{\mathrm{Total}} = H_{\mathrm{Local}} + H_{\mathrm{Remote}}
\end{equation}
Under the rotating wave approximation, the local component of the total Hamiltonian takes the form:
\begin{equation}\label{eq:2}
\begin{aligned}
H_{\text{Local}} &=\frac{\omega_q}{2} \sigma_z + \omega_c c^\dagger c + \omega_{m_{L}} {m_{L}}^\dagger {m_{L}}  \\&\quad + g_1 (\sigma_- c^{\dagger} + \sigma_+ c) + g_2 (c m_{L}^{\dagger} + c^{\dagger} m_{L})
\end{aligned}
\end{equation}
Here $c$ ($c^\dagger$) and $m_{L}$ ($m_{L}^\dagger$) represent the annihilation (creation) operators of the SC mode and the local magnon mode, with frequencies $\omega_c$ and $\omega_{m_{L}}$, respectively. Additionally, $\sigma_\pm$ are the transition operators of the SQ, with the energy level spacing frequency $\omega_q$. 
The term $g_1 (\sigma_- c^\dagger+\sigma_+ c)$ describes the interaction between SQ and SC with a coupling strength $g_1$. The magnetic dipole interaction term $g_2 (c m_{L}^\dagger+c^\dagger m_{L} )$ represents the coupling between the SC mode $c$ and the local magnon mode $m_{L}$ with strength $g_2$\cite{H1,H2,H3}. 

The remote component of the total Hamiltonian comprises four distinct contributions:
\begin{equation}\label{eq:3}
H_{\mathrm{Remote}} = H_{\mathrm{QM1}} + H_{\mathrm{WG}} + H_{\mathrm{Int}} + H_{\mathrm{QM2}}
\end{equation}
where $H_{\mathrm{QM1}}$ and $H_{\mathrm{QM2}}$ describe the Brillouin scattering processes in the two YIG spheres, respectively:
\begin{equation}\label{eq:4}
\begin{aligned}
H_{\text{QM1}} &= \omega_{a_{L}}a_{L}^\dagger a_{L} + \omega_{b_{L}}b_{L}^\dagger b_{L}  \\
&\quad+ \left(g_{L} a_{L} b_{L}^\dagger m_{L} + \sqrt{g_3}\varepsilon_1^* a_{L} e^{i \omega_{d} t} + \mathrm{H.c.}\right)
\end{aligned}
\end{equation}
\begin{equation}\label{eq:5}
\begin{aligned}
H_{\text{QM2}} &= \omega_{a_{R}}a_{R}^\dagger a_{R} + \omega_{b_{R}}b_{R}^\dagger b_{R} + \omega_{m_{R}}m_{R}^\dagger m_{R} \\
&\quad+ \left(g_{R} a_{R} b_{R}^\dagger m_{R} + \sqrt{g_6}\varepsilon_2^* a_{R} e^{i \omega_{d} t} + \mathrm{H.c.}\right)
\end{aligned}
\end{equation}
Here, $a_{L}$ ($a_{L}^\dagger$) and $b_{L}$ ($b_{L}^\dagger$) denote the annihilation (creation) operators for the WGM photons in the local YIG sphere. The operators $a_{R}$ ($a_{R}^\dagger$) and $b_{R}$ ($b_{R}^\dagger$) represent the corresponding operators in the remote YIG sphere, while $m_{R}$ ($m_{R}^\dagger$) describes the magnon mode in the remote YIG sphere.
The coherent pump fields $\varepsilon_1$ and $\varepsilon_2$ (with frequency $\omega_d$) are introduced through the waveguide with coupling rates $g_3$ and $g_6$ respectively. $g_{L}$ ($g_{R}$) denotes the magnon-photon coupling strength in the local (remote) YIG sphere\cite{M2,38,39}. 
The terms $\mathrm{H.c.}$ stand for the Hermitian conjugates of the preceding interactions.

The Hamiltonian $H_{\mathrm{WG}}$ represents the free Hamiltonian of the waveguide photons, while $H_{\mathrm{Int}}$ describes the interaction between the waveguide photon modes and the WGMs of the YIG spheres. Their explicit forms are given by:
\begin{equation}\label{eq:6}
H_{\mathrm{WG}}= \int \omega \left[ d_a^{\dagger}(\omega) d_a(\omega) + d_b^{\dagger}(\omega) d_b(\omega) \right] d\omega
\end{equation}
\begin{equation}\label{eq:7}
\begin{aligned}
H_{\mathrm{Int}} &= i \int d\omega \sqrt{\frac{g_4}{2\pi}} \left( d_a^{\dagger}(\omega) a_{L} e^{-i\omega\tau-i\omega_d t}- a_{L}^{\dagger} d_a(\omega) e^{i\omega\tau+i\omega_d t} \right) \\
&+ i \int d\omega \sqrt{\frac{g_5}{2\pi}} \left( d_a^{\dagger}(\omega) a_{R} e^{-i\omega\tau-i\omega_d t} - a_{R}^{\dagger} d_a(\omega) e^{i\omega\tau+i\omega_d t} \right) \\
&+ i \int d\omega \sqrt{\frac{g_7}{2\pi}} \left( d_b^{\dagger}(\omega) b_{L} e^{-i\omega\tau} - b_{L}^{\dagger} d_b(\omega) e^{i\omega\tau} \right)\\
&+ i \int d\omega \sqrt{\frac{g_8}{2\pi}} \left( d_b^{\dagger}(\omega) b_{R} e^{-i\omega\tau} - b_{R}^{\dagger} d_b(\omega) e^{i\omega\tau} \right) 
\end{aligned}
\end{equation}
Here, $d_{a}(\omega)$ ($d_{a}^{\dagger}(\omega)$) and $d_{b}(\omega)$ ($d_{b}^{\dagger}(\omega)$) denote the annihilation (creation) operators for photons in the waveguide mode at frequency $\omega$, which are coupled to the $a_{{L}/{R}}$ (here the notation $a_{{L}/{R}}$ denotes either the local ($a_{L}$) or remote ($a_{R}$) mode, and similarly for other subscripts with ${L}/{R}$ throughout this paper) and $b_{{L}/{R}}$ WGMs, respectively. The parameter $\tau$ corresponds to the propagation time of photons traveling through the waveguide between the local and remote YIG spheres.
The coupling strengths $g_4$ and $g_5$ ($g_7$ and $g_8$) mediate the interactions between the waveguide mode and the WGMs $a_{L}$ and $a_{R}$ ($b_{L}$ and $b_{R}$) in the local and remote YIG spheres, respectively.
The terms containing $e^{\pm i\omega_d t}$ account for the coherent driving at frequency $\omega_d$, where the phase factors $e^{\pm i\omega\tau}$ incorporate the time delay effects during photon propagation. 
The first two terms describe the interaction between the waveguide photons and the $a_{{L}/{R}}$ mode, whereas the last two terms correspond to the interaction with the $b_{{L}/{R}}$ mode.
%Sample body text. Sample body text. Sample body text. Sample body text. Sample body text. Sample body text. Sample body text. Sample body text.

\section{Two-stage protocol for distant SQ-QM2 entanglement generation}\label{sec3}
To generate distant entanglement between the SQ and the QM2, a two-stage protocol is employed. In the initialization phase, the SC is prepared in the excited state, while both SQ and QM1 are in their ground states. Entanglement between the SQ and QM1 is deterministically generated via a shortcut to adiabaticity (STA) protocol, which is implemented by precise temporal modulation of their coupling strength, driving the system into a maximally entangled Bell state. Subsequently, in the second stage, we derive the effective Hamiltonian governing the interaction between the local QM1 and the remote QM2. By leveraging this effective interaction, we implement a controlled dynamical evolution to coherently transfer the entanglement from SQ-QM1 to SQ-QM2. This two-phase approach efficiently generates distant entanglement between local and remote quantum platforms.

\subsection{Stage 1: Generation of local SQ-QM1 entanglement}
\subsubsection{Invariant-based STA}
First, the theory of invariant-based STA is briefly introduced\cite{STA1,STA2,STA3}. It is assumed that the evolution of a quantum system can be described by a time-dependent Hamiltonian $H(t)$, with an invariant $I(t)$ satisfying: $i\hbar\frac{\partial I(t)}{\partial t} - [H(t), I(t)] = 0$. 
The eigenstates of $I(t)$ can be used to represent any solution 
$\ket{\Psi(t)}$ of the time-dependent Schrödinger equation. Specifically, the general solution can be expressed as a superposition of the eigenstates of the Hamiltonian:$\ket{\Psi(t)} = \sum_{n} c_n \ket{\psi_n(t)}$, where $\ket{\psi_n(t)}$ are the eigenstates of $H(t)$, and $c_n$ are time-independent probability amplitude. According to the Lewis-Riesenfeld invariant theory, the eigenstates of $I(t)$ differ from the eigenstates of $H(t)$ by only a phase factor:
\begin{equation}\label{eq:8}
\ket{\psi_n(t)} = e^{i\alpha_n(t)} \ket{\varphi_n(t)}
\end{equation}
where $\ket{\varphi_n(t)}$ are the eigenstates of $I(t)$. The quantity $\alpha_n (t)$, referred to as the Lewis-Riesenfeld phase, is defined as:
\begin{equation}\label{eq:9}
\alpha_n(t) = \frac{1}{\hbar} \int_0^t \left\langle \varphi_n(t') \left| i\hbar \frac{\partial}{\partial t'} - H(t') \right| \varphi_n(t') \right\rangle dt'
\end{equation}

In summary, the eigenstates of $I(t)$ can be used to represent any solution of the time-dependent Schrödinger equation $\ket{\Psi(t)}$: 
\begin{equation}\label{eq:10}
\ket{\Psi(t)} = \sum_{n} c_n e^{i\alpha_n(t)} \ket{\varphi_n(t)}
\end{equation}

The above expression indicates that any solution $\ket{\Psi(t)}$ of the Schrödinger equation can be expressed as a superposition of the eigenstates $\ket{\varphi_n(t)}$ of $I(t)$. In practical applications, the initial and final states are represented by the eigenstates $\ket{\varphi_n(t)}$. The nonadiabatic evolution path constructed from the eigenstates of $I(t)$ substitutes the original adiabatic evolution path, thereby accelerating the quantum state dynamics.
\subsubsection{Generation entanglement via STA}
\begin{figure}[h]
\centering
\includegraphics[width=0.5\textwidth]{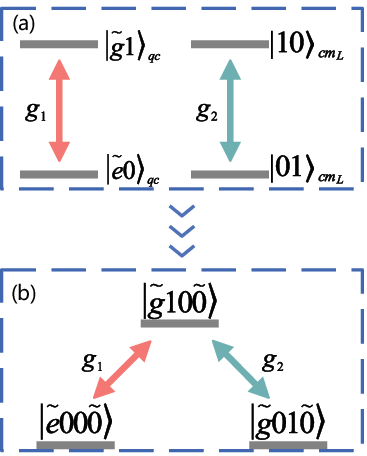}
\caption{(a)The interactions of the SC with the SQ and QM1, the state couplings $\ket{\tilde{g}1}_{qc} \leftrightarrow \ket{\tilde{e}0}_{qc}$ and $\ket{10}_{cm_{L}} \leftrightarrow \ket{01}_{cm_{L}}$. (b)Within the logical state basis: $\ket{\tilde{e}00\tilde{0}}$, $\ket{\tilde{g}10\tilde{0}}$, and $\ket{\tilde{g}01\tilde{0}}$, the coupling between $\ket{\tilde{e}00\tilde{0}}$ and $\ket{\tilde{g}01\tilde{0}}$ through the intermediate state $\ket{\tilde{g}10\tilde{0}}$.}\label{Fig3}
\end{figure}

This section presents the application of Invariant-based STA approach to generate entanglement between the local SQ and QM1\cite{Y11,Y10}. The Hamiltonian governing this process is $H_{\mathrm{Local}}$. To explicitly demonstrate the interaction, we further analyze the dynamics in the three logical states system, as in Figure~\ref{Fig3}.

The interaction Hamiltonian $H_1$ is expressed in terms of three logical states: $\ket{\tilde{e}00\tilde{0}}$, $\ket{\tilde{g}10\tilde{0}}$, and $\ket{\tilde{g}01\tilde{0}}$ as follows:
\begin{equation}\label{eq:11}
H_1 = g_1 \ket{\tilde{e}00\tilde{0}}\bra{\tilde{g}10\tilde{0}} + g_2 \ket{\tilde{g}10\tilde{0}}\bra{\tilde{g}01\tilde{0}} + \mathrm{H.c.}
\end{equation}
The Dirac bracket $\ket{\tilde{q}cm_{L}\tilde{m}_{R}}$ represents the superconducting qubit mode $\tilde{q}$, SC mode $c$, local magnon mode $m_{L}$ and remote magnon mode $\tilde{m}_{R}$, respectively. The matrix form of $H_1$ is expressed using the three logical states as follows: 
\begin{equation}\label{eq:12}
H_1 = 
\begin{pmatrix}
0 & g_1 & 0 \\
g_1^* & 0 & g_2 \\
0 & g_2^* & 0
\end{pmatrix}
\end{equation}
The invariant $I(t)$, which satisfies the equation $i\hbar\frac{\partial I(t)}{\partial t} - [H(t), I(t)] = 0$, is expressed as follows:
\begin{equation}\label{eq:13}
I(t) = \Omega_0
\begin{pmatrix}
0 & \cos\gamma \sin\theta & -i \sin\gamma \\
\cos\gamma \sin\theta & 0 & \cos\gamma \cos\theta \\
i \sin\gamma & \cos\gamma \cos\theta & 0
\end{pmatrix}
\end{equation}
The eigenstates of $I(t)$ are given by the following expression:
\begin{equation}\label{eq:14}
\ket{\varphi_0(t)} = 
\begin{pmatrix}
\cos\gamma \cos\theta \\
-i \sin\gamma \\
-\cos\gamma \sin\theta
\end{pmatrix}
\end{equation}
\begin{equation}\label{eq:15}
\ket{\varphi_{\pm}(t)} = \frac{1}{\sqrt{2}}
\begin{pmatrix}
\sin\gamma \cos\theta \pm i \sin\theta \\
i \cos\gamma \\
-\sin\gamma \sin\theta \pm i \cos\theta
\end{pmatrix}
\end{equation}
The eigenstates $\ket{\varphi_0(t)}$ and $\ket{\varphi_{\pm}(t)}$ correspond to the eigenvalues $0$ and $\pm\Omega_0$. By substituting the form of the invariant into the equation $i\hbar\frac{\partial I(t)}{\partial t} - [H(t), I(t)] = 0$, the auxiliary equations for the coupling strengths $g_1$ and $g_2$ are derived as:
\begin{equation}\label{eq:16}
\begin{aligned}
g_1 &= \dot{\theta} \cot \gamma \sin \theta + \dot{\gamma} \cos \theta, \\
g_2 &= \dot{\theta} \cot \gamma \cos \theta - \dot{\gamma} \sin \theta.
\end{aligned}
\end{equation}
Here, $\gamma(t)$ and $\theta(t)$ are time-dependent parameters, and $\dot{\gamma}$ and $\dot{\theta}$ denote their respective time derivatives. To achieve the goal of generating entanglement between the SQ and QM1, in the time interval $[0,T_1)$, the parameters $\gamma$ and $\theta$ are given by:
\begin{equation}\label{eq:17}
\gamma = \frac{\pi}{2T_1} t - \frac{\pi}{2};\quad\theta = \frac{\pi}{4}
\end{equation}
The auxiliary equations are reduced to:
\begin{equation}\label{eq:18}
g_1 = \frac{\sqrt{2}\pi}{4T_1};\quad g_2 = -\frac{\sqrt{2}\pi}{4T_1}.
\end{equation}

The objective of our study is to achieve the evolution of the quantum state from the initial state $\ket{\tilde{g}10\tilde{0}}$ to an entangled superposition state of $\ket{\tilde{e}00\tilde{0}}$ and $\ket{\tilde{g}01\tilde{0}}$. Therefore, we define the ideal initial state $\ket{\psi_{1\_\mathrm{initial}}}$ and the target final state $\ket{\psi_{1\_\mathrm{final}}}$ as follows:
\begin{equation}\label{eq:19}
\ket{\psi_{1\_\mathrm{initial}}} = \ket{\tilde{g}10\tilde{0}}
\end{equation}
\begin{equation}\label{eq:20}
\ket{\psi_{1\_\mathrm{final}}} = \frac{1}{\sqrt{2}}({\ket{\tilde{e}00\tilde{0}}+\ket{\tilde{g}01\tilde{0}}})
\end{equation}
At the initial time $t=0$, the time-independent probability amplitudes $c_{0}$ and $c_{\pm}$ are obtained. By integrating, the values of $\alpha_{n} (T_1 )$ are calculated. The results are substituted into $\ket{\Psi(T_1)} = \sum_{n} c_{n} e^{i\alpha_{n}(T_1)} \ket{\varphi_n(T_1)}$. 
In the absence of dissipation, the target state $\ket{\psi_{1\_\mathrm{final}}} = \frac{1}{\sqrt{2}}({\ket{\tilde{e}00\tilde{0}}+\ket{\tilde{g}01\tilde{0}}})$ can be obtained. By extracting vacuum states $\ket{0\tilde{0}}_{c{m_{R}}}$ of the SC mode and remote magnon mode, specifically $\frac{1}{\sqrt{2}}(\ket{\tilde{e}0}_{qm_{L}}+\ket{\tilde{g}1}_{qm_{L}})\otimes\ket{0\tilde{0}}_{c{m_{R}}}$, a maximally entangled Bell state, $\frac{1}{\sqrt{2}}(\ket{\tilde{e}0}_{qm_{L}}+\ket{\tilde{g}1}_{qm_{L}})$, is generated between the SQ and QM1.
\subsection{Stage 2: Coherent entanglement transfer to remote QM2}
\subsubsection{Effective Hamiltonian of the magnon modes}
To enhance the coupling strength between the optical modes $b_{{L}/{R}}$ and the magnon modes $m_{{L}/{R}}$, the present scheme drives the optical modes $a_{{L}/{R}}$ with pump fields $\varepsilon_{1,2}$.
Considering the coupling between the $a_{{L}/{R}}$ modes and the waveguide photon modes, and incorporating the pump fields $\varepsilon_{1,2}$ along with the intrinsic dissipation of the optical modes $a_{{L}/{R}}$, the dynamical equation for $\langle a_{{L}/{R}} \rangle$ is given by \cite{Y3,40}:
\begin{equation}\label{eq:21}
\begin{aligned}
\langle {\dot a_{{L}/{R}}} \rangle &= \left( -i\Delta_{a} - \frac{\kappa_{{L}/{R}}}{2} \right) \langle a_{{L}/{R}} \rangle \\
&\quad- ig_{{L}/{R}} \langle b_{{L}/{R}} m_{{L}/{R}}^\dagger \rangle - i \sqrt{g_{3,6}} \varepsilon_{\mathrm{eff},{{L}/{R}}}
\end{aligned}
\end{equation}
Here, $\Delta_{a}=\omega_a-\omega_d$ is the detuning between the optical mode frequency ($\omega_a = \omega_{a_{L}} = \omega_{a_{R}}$) and the drive frequency ($\omega_d$). The total decay rate $\kappa_{{L}/{R}}$ is given by: $\kappa_{{L}/{R}} = \kappa_{a_{{L}/{R}}}+g_{3,6}+ g_{4,5}$, where $\kappa_{a_{{L}/{R}}}$ is the intrinsic decay rate, and $\varepsilon_{\mathrm{eff},{{L}/{R}}}$ denotes the effective drive\cite{41}. In the derivation, we assume the waveguide is initially in the vacuum state and neglect retardation effects (Markovian approximation).
In the strong pump field limit, the higher-order optomagnetic interaction term
($- ig_{{L}/{R}} \langle b_{{L}/{R}} m_{{L}/{R}}^\dagger \rangle$) becomes negligible (given that $g_{{L}/{R}}$ is of order $\mathrm{Hz}$)\cite{42,43}. Consequently, the operator $a_{{L}/{R}}$ can be replaced by its steady-state expectation value $\langle a_{{L}/{R}} \rangle_{ss} = \frac{i \sqrt{g_{3,6}} \varepsilon_{\mathrm{eff},{{L}/{R}}}}{-i \Delta_{a} - \kappa_{{L}/{R}} / 2}$. In the rotating frame defined by $\sum_{j=L,R} \omega_{d} (a_j^\dagger a_j + b_j^\dagger b_j)$, the Hamiltonians $H_{\mathrm{QM1}}$ and $H_{\mathrm{QM2}}$ can be expressed as:
\begin{equation}\label{eq:22}
\begin{aligned}
H'_{\mathrm{QM1}} &= \Delta_{b} b_{L}^\dagger b_{L} + \omega_{m} m_{L}^\dagger m_{L} \\
&\quad+ G_{L} b_{L}^\dagger m_{L} + G_{L}^* b_{L} m_{L}^\dagger, \\
H'_{\mathrm{QM2}} &= \Delta_{b} b_{R}^\dagger b_{R} + \omega_{m} m_{R}^\dagger m_{R} \\
&\quad+ G_{R} b_{R}^\dagger m_{R} + G_{R}^* b_{R} m_{R}^\dagger.
\end{aligned}
\end{equation}
where $\Delta_{b}=\omega_b-\omega_d$ denotes the detuning between the optical mode frequency ($\omega_b=\omega_{b_{L}}=\omega_{b_{R}}$) and the drive frequency ($\omega_d$), $\omega_m=\omega_{m_{L}}=\omega_{m_{R}}$ represents the frequency of the magnon mode, and $G_{{L}/{R}}$ describes the effective coupling strength.
The effective coupling strength $G_{{L}/{R}}$ is proportional to the effective pump field $\varepsilon_{\mathrm{eff},{{L}/{R}}}$, enabling its adjustment via tuning $\varepsilon_{\mathrm{eff},{{L}/{R}}}$\cite{41,44}.
Considering the coupling between the $b_{{L}/{R}}$ modes and the waveguide photon modes, we derive the master equation describing the indirect interaction between the local and remote nodes as \cite{Y3}:
\begin{equation}\label{eq:23}
\begin{aligned}
\dot{\rho_b} &= -i[H'_{\mathrm{node}}, \rho_b] + g_{7} \mathcal{L}[b_{L}]\rho_b + g_{8} \mathcal{L}[b_{R}]\rho_b \\
&\quad- \sqrt{g_{7}g_{8}} \left( e^{-ikx} [\rho_b b_{L}^\dagger, b_{R}] + e^{ikx} [b_{R}^\dagger, b_{L} \rho_b] \right)
\end{aligned}
\end{equation}
where $H'_{\mathrm{node}} = H'_{\mathrm{QM1}} + H'_{\mathrm{QM2}}$, $\mathcal{L}[\Theta]\rho = \Theta \rho \Theta^\dagger - \frac{1}{2} \{ \Theta^\dagger \Theta, \rho \}$ denotes the Lindblad dissipative superoperator, and $e^{-ikx}$ represents the phase delay accumulated by photons propagating from the local node to the remote node, where $k$ denotes the wavenumber in the waveguide and $x$ is the separation between the nodes. The first term governs the coherent evolution of the system. The second and third terms describe energy dissipation at individual nodes, while the fourth term accounts for photon transmission between nodes.

To derive an effective magnon Hamiltonian, we eliminate the optical modes $b_{{L}/{R}}$ under the assumption that the timescale for the optical modes $b_{{L}/{R}}$ to reach their steady state is much shorter than the timescale of the interaction between the optical and magnon modes.
By employing the Nakajima-Zwanzig projection operator technique, we derive the Lindblad-form master equation for the magnon density matrix $\mu = \text{Tr}_b[\rho]$\cite{Y3,45}:
\begin{equation}\label{eq:24}
\begin{aligned}
\dot{\mu}&= -i[H_{\mathrm{Remote}}^{\mathrm{eff}}, \mu] \\
&\quad+ \widetilde{S}_{LL} \left( 2 m_{L} \mu m_{L}^\dagger - m_{L}^\dagger m_{L} \mu - \mu m_{L}^\dagger m_{L} \right) \\
&\quad+ \widetilde{S}_{LR} \left( 2 m_{R} \mu m_{L}^\dagger - m_{L}^\dagger m_{R} \mu - \mu m_{L}^\dagger m_{R} \right) \\
&\quad+ \widetilde{S}_{RL} \left( 2 m_{L} \mu m_{R}^\dagger - m_{R}^\dagger m_{L} \mu - \mu m_{R}^\dagger m_{L} \right) \\
&\quad+ \widetilde{S}_{RR} \left( 2 m_{R} \mu m_{R}^\dagger - m_{R}^\dagger m_{R} \mu - \mu m_{R}^\dagger m_{R} \right)
\end{aligned}
\end{equation}
where $\widetilde{S}_{jk} = \frac{G_j^* G_k}{2}\left[ S_{jk}(\omega_{m}) + S_{kj}^*(\omega_{m}) \right] $ with $S_{jk}(\omega) = \int_{0}^{\infty} \langle c_j(\tau) c_k^\dagger(0) \rangle e^{i\omega \tau} d\tau \ (j,k={L},{R})$. 
Assuming $\omega_{m_{{L}/{R}}} = \Delta_{b} = \omega_m$, $g_{7} = g_{8} = g$, $kx=2n\pi$ ($n$ is an integer) and the effective coupling $G_{{L}/{R}}$ as real, we can calculate the coefficients $S_{jj}(\omega_m) = \frac{2}{g} \ (j = {L},{R}), \ S_{RL}(\omega_m) = -\frac{4}{g}, \ S_{LR}(\omega_m) = 0$. The effective Hamiltonian $H_{\mathrm{Remote}}^{\mathrm{eff}}$ can be written in the rotating frame $\omega_m m_{L}^\dagger m_{L} + \omega_m m_{R}^\dagger m_{R}$ as:
\begin{equation}\label{eq:25}
H_{\mathrm{Remote}}^{\mathrm{eff}} = g_{\mathrm{eff}}m_{R}^\dagger m_{L} + g_{\mathrm{eff}}^* m_{L}^\dagger m_{R}
\end{equation}
where $g_{\mathrm{eff}}=\frac{2i}{g} G_{L} G_{R}$ represents the effective coupling strength between the magnons.
\subsubsection{Dynamical evolution for entanglement transfer}
\begin{figure}[h]
\centering
\includegraphics[width=0.7\textwidth]{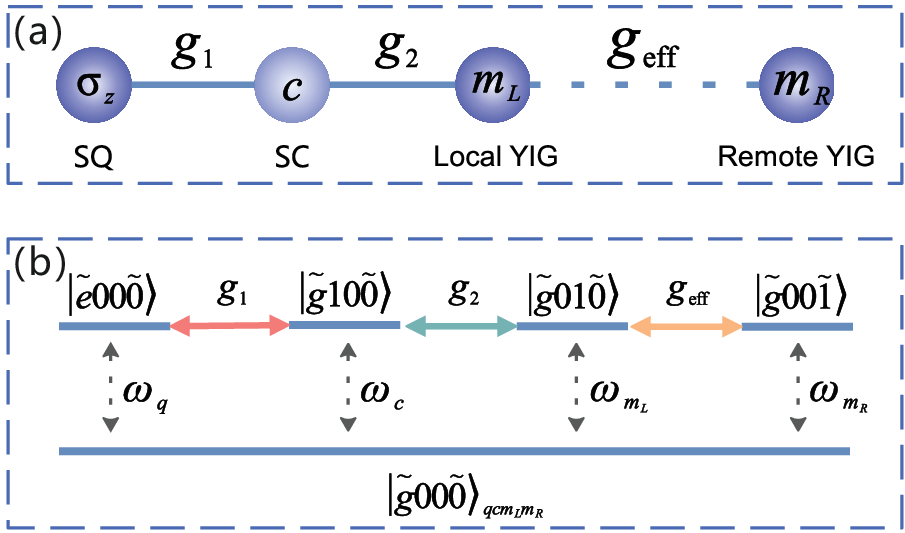}
\caption{(a)The simplified coupling diagram. (b)We use five logical states to describe the coupling between the four subsystems, where $g_{\mathrm{eff}}$ represents the effective coupling strength between $m_{L}$ and $m_{R}$. $\ket{\tilde{g}00\tilde{0}}_{qcm_{L}m_{R}}$ represents the state in which all subsystems are in their ground states.}\label{Fig4}
\end{figure}

Based on the effective Hamiltonian between the magnon modes derived in the previous section, this section applies dynamical evolution to transfer the entanglement from the SQ-QM1 systems (obtained after the first stage) to the SQ-QM2 systems. To clearly illustrate the interactions between the subsystems, we simplify the coupling diagram as shown in Figure~\ref{Fig4}.

\begin{figure}[h]
\centering
\includegraphics[width=0.7\textwidth]{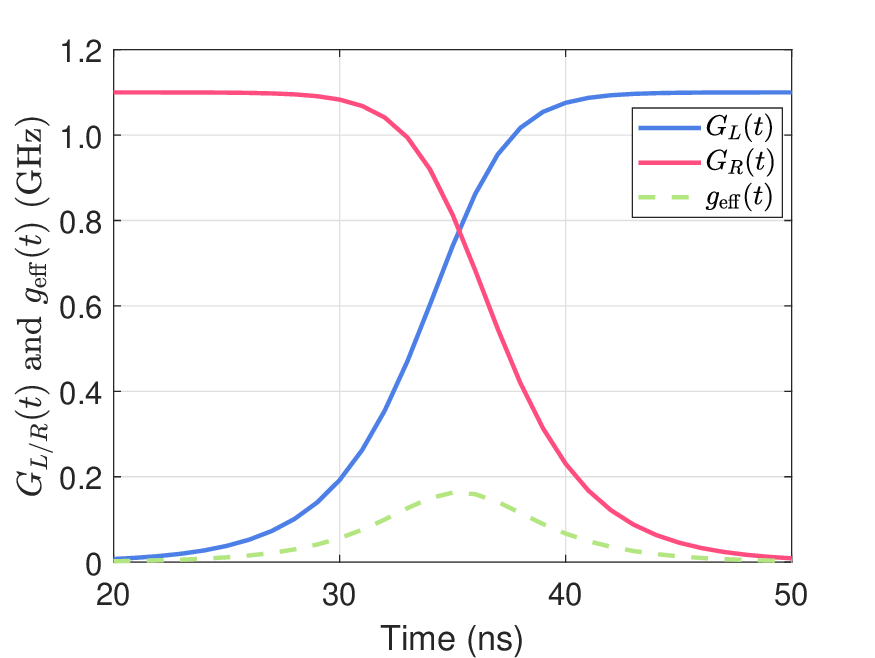}
\caption{Designed temporal profiles of the effective coupling strengths $G_{{L}/{R}}(t)$ and $g_{\mathrm{eff}}(t)$.}\label{Fig5}
\end{figure}

Following the generation of local SQ-QM1 entanglement yielding the state $\ket{\psi_{1\_\mathrm{final}}} = \frac{1}{\sqrt{2}}\left(\ket{\tilde{e}00\tilde{0}} + \ket{\tilde{g}01\tilde{0}}\right)$, we implement the entanglement transfer protocol during the interval $[T_1, T_2)$. At $t = T_1$, we eliminate the local couplings by setting $g_1 = g_2 = 0$ while activating the effective coupling $g_{\mathrm{eff}}$. To achieve efficient entanglement transfer from the SQ-QM1 to the SQ-QM2, we engineer the effective coupling strength following the method described in~\cite{Y3}. The coupling parameters $G_{L}$ and $G_{R}$ are designed as:
\begin{equation}\label{eq:26}
\begin{aligned}
 G_{L}(t) &= \Omega \sqrt{\frac{1-\left\{\frac{1 - \exp\left[\frac{4\Omega^2}{g} \left(t-T_1\right) - 10 \right]}{1 + \exp\left[ \frac{4\Omega^2}{g} \left(t-T_1\right) - 10 \right]}\right\}}{2}}  \\
G_{R}(t) &= \Omega \sqrt{\frac{1+\left\{\frac{1 - \exp\left[\frac{4\Omega^2}{g} \left(t-T_1\right) - 10 \right]}{1 + \exp\left[ \frac{4\Omega^2}{g} \left(t-T_1\right) - 10 \right]}\right\}}{2}} 
\end{aligned}
\end{equation}
where $\Omega = \sqrt{G_{L}^2(t)+G_{R}^2(t)} = 0.15g$ is held constant. 
Figure~\ref{Fig5} shows the temporal profiles of $G_{L}(t)$ and $G_{R}(t)$. We precisely control these couplings to maintain the effective Hamiltonian equation~(\ref{eq:25}), which enables coherent entanglement transfer, while suppressing unwanted transitions. In the absence of dissipation, the system evolves under the influence of $g_{\mathrm{eff}}(t)$ to the state:
\begin{equation}\label{eq:27}
\ket{\psi_{\mathrm{target}}} = \frac{1}{\sqrt{2}}\left(\ket{\tilde{e}00\tilde{0}} + \ket{\tilde{g}00\tilde{1}}\right)
\end{equation}
By extracting vacuum states $\ket{00}_{c{m_{L}}}$ of the SC mode and local magnon mode, specifically $\frac{1}{\sqrt{2}}(\ket{\tilde{e}\tilde{0}}_{qm_{R}}+\ket{\tilde{g}\tilde{1}}_{qm_{R}})\otimes\ket{00}_{c{m_{L}}}$, a maximally entangled Bell state, $\frac{1}{\sqrt{2}}(\ket{\tilde{e}\tilde{0}}_{qm_{R}}+\ket{\tilde{g}\tilde{1}}_{qm_{R}})$, is generated between the SQ and QM2.

This final state confirms complete transfer of entanglement to SQ-QM2 systems. The protocol requires precise temporal control of coupling strengths. Ultimately, this enables deterministic generation of entanglement between local (SQ) and remote (QM2) quantum systems. The following sections will consider the impact of dissipation on the entire process.
\section{Numerical simulation}\label{sec4}
To demonstrate the advantages of the proposed two-stage protocol for generating high-fidelity entanglement between the local (SQ) and remote (QM2) quantum systems, we perform detailed numerical simulations. The protocol consists of two distinct phases: (i) evolution under a coupling strength engineered via the invariant-based STA method, followed by (ii) evolution under a predetermined effective coupling strength. Crucially, our simulations account for decoherence effects in both phases to evaluate the protocol's robustness under realistic conditions.
We quantitatively characterize the resulting SQ-QM2 entangled state through two key metrics: the state fidelity and the Negativity, a well-established entanglement measure. Furthermore, we systematically analyze the system's performance across various decoherence regimes, demonstrating the practical advantages of this entanglement generation scheme.

\subsection{Entanglement generation between local and remote systems}
Based on the simplified coupling scheme depicted in Figure~\ref{Fig4}, we construct a composite system comprising four two-dimensional subsystems in a tensor product space. During the initial stage $\left[0,T_1\right)$, we generate SQ-QM1 entanglement by implementing the coupling strengths $g_1$ and $g_2$, which are designed according to equation~(\ref{eq:18}). Notably, the effective coupling strength is maintained at zero throughout this phase. The subsequent stage $\left[T_1,T_2\right)$ achieves entanglement transfer from the SQ-QM1 systems to SQ-QM2 systems. Here, we set $g_1 = g_2 = 0$, enabling the entanglement transfer to proceed solely under the influence of the effective coupling $g_{\mathrm{eff}}(t)$.
To fully characterize the system's evolution under these controlled couplings, we utilize the Lindblad master equation:
\begin{equation}\label{eq:28}
\begin{aligned}
\dot{\rho} &= i[\rho, H'_{\mathrm{Total}}] \\
&\quad+ \widetilde{S}_{LL} \left( 2 m_{L} \mu m_{L}^\dagger - m_{L}^\dagger m_{L} \mu - \mu m_{L}^\dagger m_{L} \right) \\
&\quad+ \widetilde{S}_{LR} \left( 2 m_{R} \mu m_{L}^\dagger - m_{L}^\dagger m_{R} \mu - \mu m_{L}^\dagger m_{R} \right) \\
&\quad+ \widetilde{S}_{RL} \left( 2 m_{L} \mu m_{R}^\dagger - m_{R}^\dagger m_{L} \mu - \mu m_{R}^\dagger m_{L} \right) \\
&\quad+ \widetilde{S}_{RR} \left( 2 m_{R} \mu m_{R}^\dagger - m_{R}^\dagger m_{R} \mu - \mu m_{R}^\dagger m_{R} \right)\\
&\quad+ \gamma_q \mathcal{L}[\sigma_-]\rho + \frac{\gamma_\phi}{2} \mathcal{L}[\sigma_z]\rho  \\
&\quad + \kappa_c (\bar{n}_c + 1) \mathcal{L}[c]\rho + \kappa_c \bar{n}_c \mathcal{L}[c^\dagger]\rho \\
&\quad + \kappa_{m_{L}} (\bar{n}_{m_{L}} + 1) \mathcal{L}[m_{L}]\rho + \kappa_{m_{L}} \bar{n}_{m_{L}} \mathcal{L}[m_{L}^\dagger]\rho \\
&\quad + \kappa_{m_{R}} (\bar{n}_{m_{R}} + 1) \mathcal{L}[m_{R}]\rho + \kappa_{m_{R}} \bar{n}_{m_{R}} \mathcal{L}[m_{R}^\dagger]\rho 
\end{aligned}
\end{equation}
Here, $\gamma_q$ and $\gamma_\phi$ denote relaxation and dephasing rates of the superconducting qubit, respectively, while $\kappa_c$, $\kappa_{m_{L}}$, and $\kappa_{m_{R}}$ correspond to the decay rates of the cavity mode, the local magnon mode, and the remote magnon mode, respectively\cite{A45}. The average thermal photon and magnon numbers, $\bar{n}_c$, $\bar{n}_{m_{L}}$, and $\bar{n}_{m_{R}}$ follow the Bose-Einstein distribution:
$\bar{n}_i = \left( \exp\left[ \frac{\hbar \omega_i}{\mathrm{k_B} T_{\mathrm{th}}} \right] - 1 \right)^{-1}$
where $i \in \{c, m_{L}, m_{R}\}$, $\hbar$ is the reduced Planck constant, $\omega_i$ is the characteristic frequency of the respective mode, $\mathrm{k_B}$ is the Boltzmann constant, and $T_{\mathrm{th}}$ is the thermal equilibrium temperature.

\begin{figure}[h]
\centering
\includegraphics[width=0.7\textwidth]{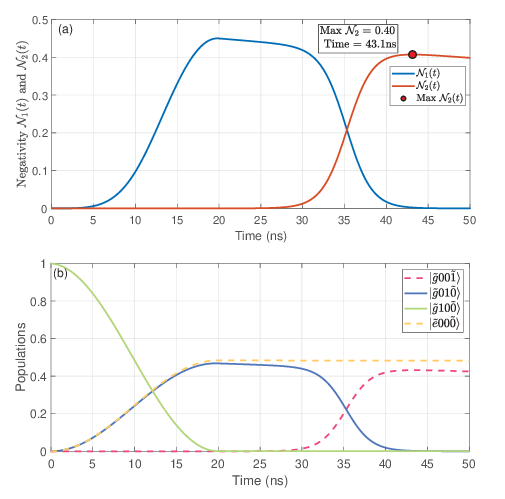}
\caption{Remote entanglement generation between SQ and QM2 via a two-stage protocol. (a)Time evolution of two-body entanglement $\mathcal{N}_1(t)$ and $\mathcal{N}_2(t)$. (b)Time evolution of various populations across two stages. During the interval $\left[0,T_1\right)$, the relaxation and dephasing rates used in the simulation are $\gamma_q/2\pi=0.01\mathrm{MHz}$ and $\gamma_{\phi}/2\pi=0.1\mathrm{MHz}$, respectively. The system frequencies are set to $\omega_c/2\pi=10\mathrm{GHz}$, $\omega_{m_{L}}/2\pi=5\mathrm{GHz}$. The thermal equilibrium temperature is $T_\mathrm{th}=50\mathrm{mK}$, and the dissipation of the subsystems are $\kappa_c=\kappa_{m_{L}}=0.5\mathrm{MHz}$. During the interval $\left[T_1,T_2\right)$, The dissipation of the subsystems used in the simulation are $\kappa_c=\kappa_{m_{L}}=\kappa_{m_{R}}=0.5\mathrm{MHz}$. The system frequencies are set to $\omega_{m_{L}}/2\pi=\omega_{m_{R}}/2\pi=5\mathrm{GHz}$. The thermal equilibrium temperature is $T_\mathrm{th}=50\mathrm{mK}$.}\label{Fig6}
\end{figure}

The system is initialized in the pure state $\rho(0) = \ket{\tilde{g}10\tilde{0}}\bra{\tilde{g}10\tilde{0}}$, with excitation exclusively localized in the superconducting circuit resonator.
We numerically solve the Lindblad master equation to obtain the time-dependent density matrix $\rho(t)$, enabling analysis of quantum correlations. Key metrics include: (i) the fidelity $F = \braket{\psi_{\mathrm{target}}|\rho(T_{2})|\psi_{\mathrm{target}}}$ quantifying overlap with the target state $\ket{\psi_{\mathrm{target}}}$, and (ii) bipartite entanglement characterized via negativity, specifically $\mathcal{N}_1$ between the SQ and the QM1, and $\mathcal{N}_2$ between the SQ and the QM2.
The corresponding entanglement dynamics $\mathcal{N}_1(t)$ and $\mathcal{N}_2(t)$ are presented in \mbox{Figure~\ref{Fig6}(a)}, showing their evolution throughout the process.
\mbox{Figure~\ref{Fig6}(b)} displays the population dynamics of the basis states $\ket{\tilde{e}00\tilde{0}}$, $\ket{\tilde{g}10\tilde{0}}$, $\ket{\tilde{g}01\tilde{0}}$, and $\ket{\tilde{g}00\tilde{1}}$, revealing coherent excitation transfer.

As demonstrated in Figure~\ref{Fig6}, our scheme generates high-fidelity ($F = 0.90$) distant entanglement between the local SQ and remote QM2, achieving a negativity of $\mathcal{N}_2(t) = 0.40$ under realistic decoherence.

\subsection{Robustness of the distant entanglement generation scheme}
Dissipation is an inherent challenge in quantum information processing, particularly in quantum communication and computing, where robust entanglement generation is essential.
In this section, we analyze the dependence of the fidelity $F$ and the maximum entanglement measure $\mathcal{N}_2(t)$ on the dissipation coefficients to assess the robustness of the proposed scheme.

\begin{figure}[h]
\centering
\includegraphics[width=0.9\textwidth]{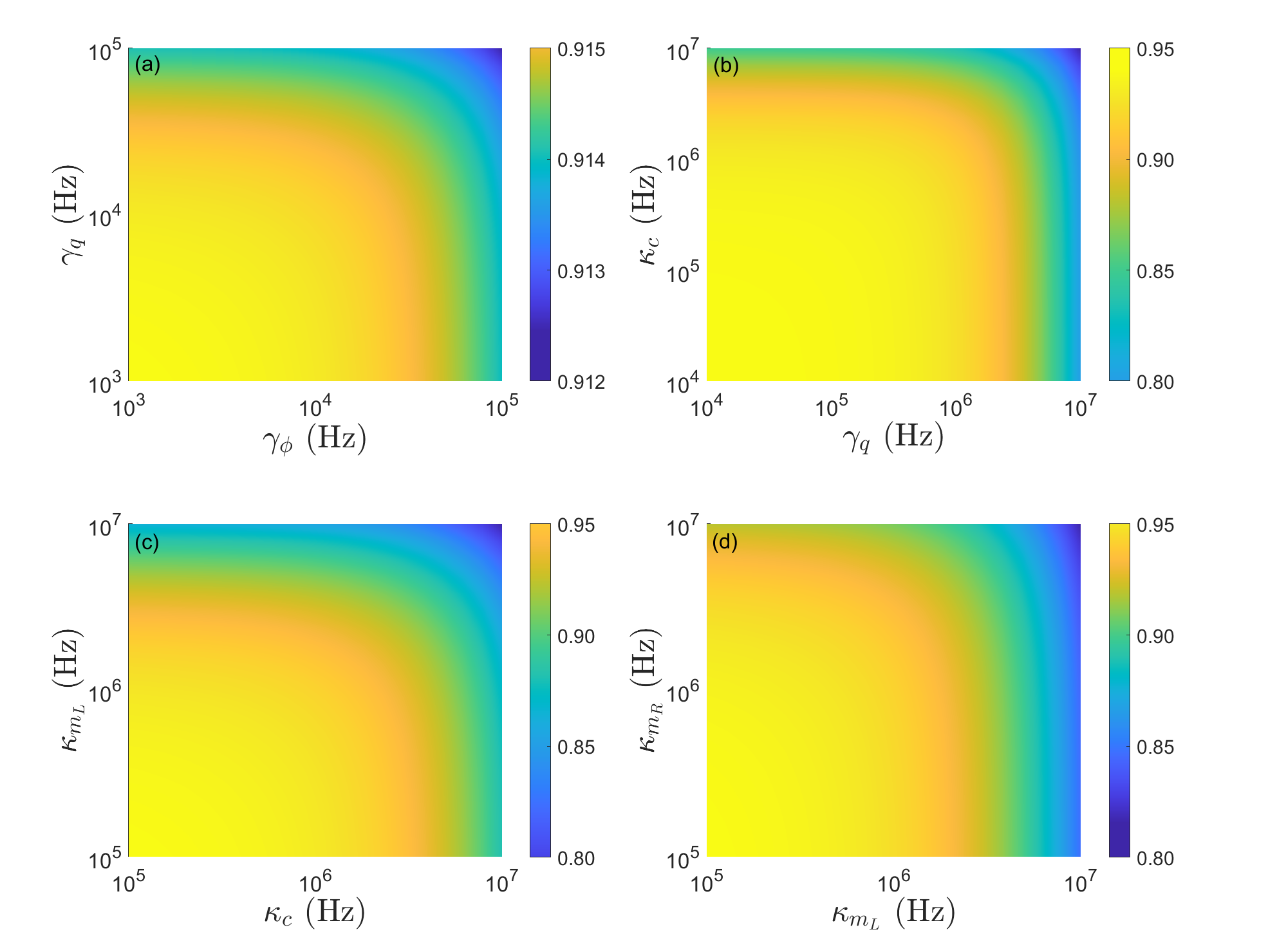}
\caption{Dependence of fidelity $F$ on relaxation rate $\gamma_q$, dephasing rate $\gamma_{\phi}$ and decay rates $\kappa_c$, $\kappa_{m_{L}}$, and $\kappa_{m_{R}}$. (a)Impact of relaxation rate $\gamma_q$ and dephasing rate $\gamma_{\phi}$ on fidelity $F$. (b)Impact of decay rate $\kappa_c$ and relaxation rate $\gamma_q$ on fidelity $F$. (c)Impact of decay rate $\kappa_c$ and dephasing rate $\kappa_{m_{L}}$ on fidelity $F$. (d)Impact of decay rate $\kappa_{m_{R}}$ and dephasing rate $\kappa_{m_{L}}$ on fidelity $F$.}\label{Fig7}
\end{figure}

Figure~\ref{Fig7} illustrates, via two-dimensional density plots, the impact of various dissipative mechanisms on the fidelity $F$ throughout the entire process of generating distant entanglement. It elucidates the predominant dissipative mechanisms affecting fidelity. As illustrated in \mbox{Figure~\ref{Fig7}(a)}, the dephasing rate $\gamma_{\phi}$ causes a larger reduction in $F$ than relaxation rate $\gamma_q$, because $\gamma_{\phi}$ directly suppresses the phase coherence of the quantum state, which is a crucial factor for the fidelity between the quantum state and the ideal state. \mbox{Figure~\ref{Fig7}(b)} reveals that the influence of relaxation rate $\gamma_q$ on fidelity surpasses that of the decay rate $\kappa_{c}$. \mbox{Figure~\ref{Fig7}(c)} shows comparable but distinguishable impacts between the cavity decay rate $\kappa_c$ and $\kappa_{m_{L}}$, with the latter exhibiting marginally stronger fidelity degradation. As further clarified in \mbox{Figure~\ref{Fig7}(d)}, the decay rate $\kappa_{m_{L}}$ demonstrates significantly stronger fidelity suppression than the remote magnon decay rate $\kappa_{m_{R}}$.
This dissipative relation demonstrates that the decay rates of the directly coupled bosonic modes--specifically the cavity mode $c$ and local magnon mode $m_{L}$--exhibit stronger impact on fidelity than the decay rate of the single indirectly coupled bosonic mode, the remote magnon mode $m_{R}$.

\begin{figure}[h]
\centering
\includegraphics[width=0.9\textwidth]{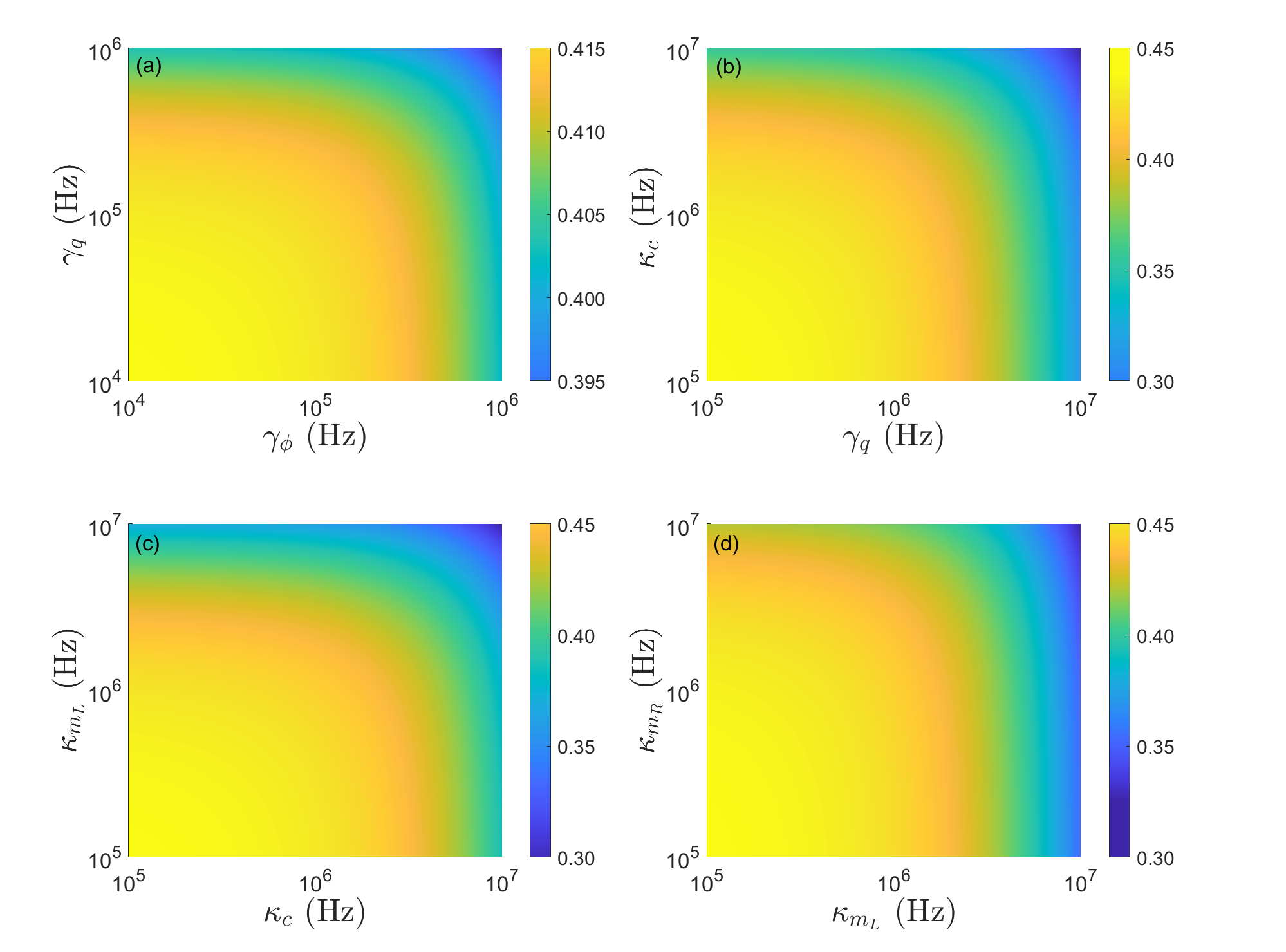}
\caption{Dependence of entanglement measure $\mathcal{N}_2$ on relaxation rate $\gamma_q$, dephasing rate $\gamma_{\phi}$ and decay rates $\kappa_c$, $\kappa_{m_{L}}$, and $\kappa_{m_{R}}$. (a)Impact of relaxation rate $\gamma_q$ and dephasing rate $\gamma_{\phi}$ on entanglement measure $\mathcal{N}_2$. (b)Impact of decay rate $\kappa_c$ and relaxation rate $\gamma_q$ on entanglement measure $\mathcal{N}_2$. (c)Impact of decay rate $\kappa_c$ and dephasing rate $\kappa_{m_{L}}$ on entanglement measure $\mathcal{N}_2$. (d)Impact of decay rate $\kappa_{m_{R}}$ and dephasing rate $\kappa_{m_{L}}$ on entanglement measure $\mathcal{N}_2$.}\label{Fig8}
\end{figure}

Figure~\ref{Fig8} presents density plots characterizing the influence of distinct dissipative channels on the entanglement measure $\mathcal{N}_2$ during the entanglement generation dynamics. Analysis of these plots reveals that the relative impact of different decay rates on $\mathcal{N}_2$ precisely mirrors the pattern observed for the fidelity $F$ in Figure~\ref{Fig7}. Specifically, \mbox{Figure~\ref{Fig8}(a)} demonstrates that the dephasing rate $\gamma_{\phi}$ induces significantly greater degradation of $\mathcal{N}_2$ than the relaxation rate $\gamma_q$. Furthermore, \mbox{Figure~\ref{Fig8}(b)} shows that $\gamma_q$ predominates over the cavity decay rate $\kappa_c$ in suppressing the entanglement measure. \mbox{Figure~\ref{Fig8}(c)} reveals a slightly stronger detrimental effect on $\mathcal{N}_2$ from the local magnon decay rate $\kappa_{m_{L}}$ compared to $\kappa_c$. Finally, \mbox{Figure~\ref{Fig8}(d)} establishes that $\kappa_{m_{L}}$ exerts a markedly stronger inhibitory effect on the entanglement measure than the remote magnon decay rate $\kappa_{m_{R}}$.

These results indicate that, within the studied parameter regime, relaxation and dephasing processes dominate the limitations on both fidelity $F$ and entanglement measure $\mathcal{N}_2$, whereas the decay rates of individual bosonic modes exhibit relatively weaker effects on these two quantities.
Crucially, the protocol successfully achieves distant entanglement generation between the local SQ and remote QM2, demonstrating resilience against these dominant decoherence channels while maintaining robust dynamical performance. This establishes the protocol's strong potential for practical implementation in large-scale quantum networks.

\section{Conclusions}\label{sec5}

We propose an efficient two-stage entanglement generation scheme for distant quantum systems within a magnon-mediated hybrid physical system. The protocol exploits a superconducting circuit resonator (SC) mediating interactions between a local superconducting qubit (SQ) and quantum magnonic system (QM1), with QM1 waveguide-coupled to a remote quantum magnonic system (QM2). This protocol establishes a deterministic entanglement generation framework: Bell state entanglement is first generated between SQ and QM1 via shortcut-to-adiabaticity (STA) protocol, followed by coherent transfer to QM2 through engineered effective Hamiltonian dynamics.
This scheme distinctively exploits magnons, which act both as functional qubits and mediators, thus significantly simplifying the overall system design.
With continued advances in experimental materials and technologies, the coupling strength is expected to increase further\cite{B44}, leading to enhanced fidelity and Negativity. These improvements will enable promising applications of this scheme, establishing a reliable pathway toward efficient long-distance quantum communication. Moreover, the proposed scheme can be readily extended to multi-node systems, demonstrating its potential for scalable quantum network construction.

\section*{Declarations}

\begin{itemize}
\item Funding

The authors declare that no funds, grants, or other support were received during the preparation of this manuscript.

\item Competing interests

The authors have no relevant financial or nonfinancial interests to disclose.

\item Ethics approval and consent to participate

Not applicable.

\item Consent for publication

All authors have read and approved the final version of the manuscript, confirm that the content of the manuscript is accurate and complete, and agree to its submission to \textit{International Journal of Theoretical Physics}.

\item Data availability

Data sharing is not applicable to this article as no datasets were generated or analyzed during the current study.

\item Materials availability

Not applicable.

\item Code availability

Not applicable.

\item Author contribution

All authors contributed to the study conception and design. The entirety of the methodological work, including the model construction, the research scheme, and numerical simulations, was performed by Guosen Liu and Pei Pei. The first draft of the manuscript was written by Guosen Liu, and all authors commented on previous versions of the manuscript. All authors read and approved the final manuscript.

\end{itemize}
\bibliography{sn-bibliography}
\end{document}